\documentclass[pra,aps,twocolumn]{revtex4}

\usepackage{amsmath}
\usepackage{bm}
\usepackage{graphicx}

\begin{document}

\title{Modulation of a quantized vortex street with a vibrating obstacle}

\author{Hiroki Saito}
\affiliation{Department of Engineering Science, University of
Electro-Communications, Tokyo 182-8585, Japan}

\author{Kenta Tazaki}
\affiliation{Department of Engineering Science, University of
Electro-Communications, Tokyo 182-8585, Japan}

\author{Tomohiko Aioi}
\affiliation{Department of Engineering Science, University of
Electro-Communications, Tokyo 182-8585, Japan}

\date{\today}

\begin{abstract}
Dynamics of a superfluid flow past an obstacle are investigated by solving
the Gross--Pitaevskii equation numerically.
For appropriate velocity and size of the obstacle, quantized vortices
are periodically generated in the wake, which form a B\'enard--von
K\'arm\'an vortex street.
It is found that vibration of an obstacle modulates the vortex street,
breaking a symmetry.
The parameter dependence of the modulation dynamics of vortices is
investigated.
\end{abstract}

\maketitle


\section{Introduction}

When a cylindrical obstacle is moved in a viscous fluid with an
appropriate velocity, vortices are periodically shed in the wake, forming
a long-lived vortex street.
This phenomenon was experimentally studied by B\'enard~\cite{Benard} and
the stability of the vortex street was theoretically investigated by von
K\'arm\'an~\cite{Karman,Lamb} in the early twentieth century.
The B\'enard--von K\'arm\'an (BvK) vortex streets are observed in a wide
scale throughout nature: from a bathtub to the atmosphere on a planet.

The behaviors of a classical viscous fluid past an obstacle are
characterized by the Reynolds number, Re, which is proportional to the
size and velocity of the obstacle and inversely proportional to the
kinematic viscosity.
A stationary flow is stable for a small value of Re.
When Re exceeds a critical value, an instability sets in and vortices are
shed in the wake, forming the BvK vortex street.
For a very large value of Re, the flow behind the obstacle becomes
turbulence.
Thus, in a classical fluid, the viscosity, which is included in Re, plays
an important role for the vortex street generation.

In contrast, Sasaki {\it et al.}~\cite{Sasaki} theoretically predicted
that a BvK-like vortex street can also be generated in superfluids, in
which viscosity is absent and a vortex is quantized.
This prediction has been made by numerical simulations of the
Gross--Pitaevskii (GP) equation, and therefore the phenomenon can be
observed in the systems that are described by the GP equation, such as an
atomic-gas Bose--Einstein condensate~\cite{Neely} and an exciton--polariton
superfluid in a semiconductor microcavity~\cite{Saito,Carusotto12}.
Theoretical studies on the dynamics of superfluids behind a moving
obstacle have been performed by many
researchers~\cite{Frisch,Jackson,Wini99,Nore00,Sties,Huepe,Aftalion}.

In the present paper, we extend the study in Ref.~\cite{Sasaki} to the
case in which an obstacle moves along a sinusoidal trajectory.
In classical fluids, it has been found that such a vibrating cylinder in a
flow generates various types of vortex streets, in which the symmetry
between two rows of vortices is broken~\cite{Griffin,Williamson}.
We will show that such symmetry breaking occurs also for the BvK vortex
street in a superfluid.
When the oscillation frequency and the vortex shedding frequency satisfy
some resonance condition, the vortex street is shown to be modulated and
become asymmetric.

This paper is organized as follows.
Section~\ref{s:form} formulates the problem.
Section~\ref{s:str} briefly reviews the case of straight motion of an
obstacle~\cite{Sasaki} and Sec.~\ref{s:sin} shows the case of sinusoidal
motion, which is the main result of the paper.
Section~\ref{s:conc} gives conclusions to the study.

\section{Formulation of the problem}
\label{s:form}

We consider the two-dimensional GP equation given by
\begin{equation} \label{GP}
i \hbar \frac{\partial\psi}{\partial t} = -\frac{\hbar^2}{2m} \nabla^2
\psi + V \psi + g |\psi|^2 \psi,
\end{equation}
where $\psi(x, y, t)$ is the macroscopic wave function, $V(x, y, t)$ is an
obstacle potential, and $g$ is an interaction coefficient.
The obstacle potential is assumed to have a Gaussian shape as
\begin{equation} \label{gauss}
V(\bm{r}, t) = V_0 \exp\{-[\bm{r} - \bm{r}_{\rm pot}(t)]^2 / d^2\},
\end{equation}
where $V_0$ is the peak intensity, $\bm{r}_{\rm pot}(t)$ is the position
of the center, and $d$ is the $1/e$ width.
For an atomic Bose--Einstein condensate, such a potential can be produced
by a blue-detuned ($V_0 > 0$) or red-detuned ($V_0 < 0$) Gaussian laser
beam.
In the present study, we consider the case of $V_0 > 0$.
The vortex generation dynamics for $V_0 < 0$ has been studied in
Ref.~\cite{Aioi}.
The potential is moved sinusoidally as
\begin{equation} \label{rpot}
\bm{r}_{\rm pot}(t) = -v t \hat{\bm{x}} + a \sin(\omega t) \hat{\bm{y}},
\end{equation}
where $\hat{\bm{x}}$ and $\hat{\bm{y}}$ are the unit vectors, $v > 0$ is
the velocity in the $-x$ direction, and $a$ and $\omega$ are the amplitude
and frequency of the oscillation.

We numerically solve Eq.~(\ref{GP}) using the pseudospectral
method~\cite{Recipes}.
The initial state is the ground state of Eq.~(\ref{GP}) with the initial
position of the Gaussian potential, which is obtained by the
imaginary-time propagation method.
The norm of the wave function, $\int |\psi|^2 d\bm{r}$, is chosen such
that the density $n_0$ far from the obstacle potential becomes a desired
value.
To break the exact numerical symmetry, small white noise is added to the
initial state.
We take a large enough space and the boundary condition does not affect
the result.

The characteristic length scale of the system is the healing length
defined by $\xi = \hbar / (m g n_0)^{1/2}$.
The sound velocity for the uniform density $n_0$ is $v_{\rm s} = (g n_0 /
m)^{1/2}$.
The characteristic time scale and energy are defined as $\tau = \hbar / (g
n_0)$ and $\mu = g n_0$, respectively.
For a typical experimental system of an atomic Bose--Einstein condensate,
$\xi \sim 0.1$ $\mu{\rm m}$, $v_{\rm s} \sim 1$ ${\rm mm} / {\rm s}$, and
$\tau \sim 0.1$ ms.
For an exciton--polariton superfluid in a semiconductor microcavity, $\xi
\sim 1$ $\mu{\rm m}$, $v_{\rm s} \sim 10^6$ ${\rm m} / {\rm s}$, and $\tau
\sim 1$ ps~\cite{Carusotto12}.

The GP equation in Eq.~(\ref{GP}) has a scaling property.
Normalizing the length, time, and density by $\xi$, $\tau$, and $n_0$,
respectively, we can eliminate the interaction coefficient $g$ from the
equation as
\begin{equation}
i \frac{\partial\tilde\psi}{\partial \tilde t} = -\frac{1}{2}
\tilde\nabla^2 \tilde\psi + \tilde V \tilde\psi + |\tilde\psi|^2
\tilde\psi,
\end{equation}
where tildes are put to the normalized quantities.
The independent parameters are therefore $V_0$, $d$, $v$, $a$, and
$\omega$.
When $V_0$ is much larger than the chemical potential $\mu$, the Gaussian
potential in Eq.~(\ref{gauss}) is almost equivalent to a circular hard-wall
potential with an effective radius $R$, given by
\begin{equation}
V_0 e^{-R^2/d^2} \simeq \mu.
\end{equation}
The height $V_0$ and $1/e$ width $d$ of the potential are thus not
independent parameters for $V_0 \gg \mu$.

\section{Numerical results}
\label{s:num}

\subsection{Straight motion of an obstacle}
\label{s:str}

\begin{widetext}
\begin{figure*}[tb]
\centerline{%
\includegraphics[width=12cm]{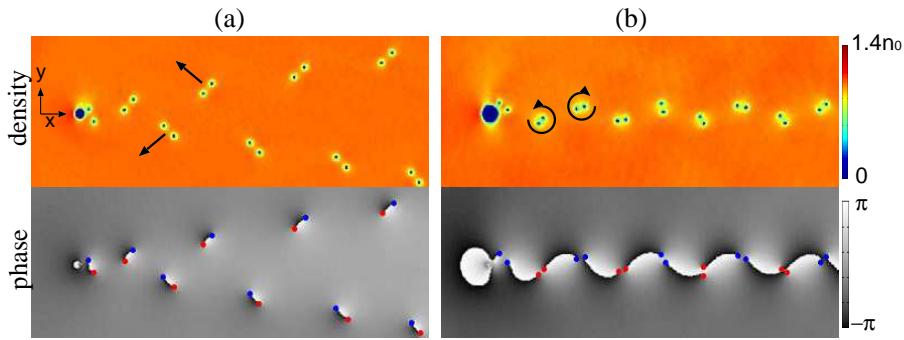}
}
\vskip6pt
\caption{Snapshots of the density ($|\psi|^2$) and phase (arg $\psi$)
profiles for the straight motion of the obstacle, $a = 0$.
(a) $v = 0.48 v_{\rm s}$, $d = 1.26 \xi$, and $V_0 = 100 \mu$.
(b) $v = 0.44 v_{\rm s}$, $d = 2.24 \xi$, and $V_0 = 100 \mu$.
The arrows in (a) indicate directions of propagation of vortex dipoles.
The arrows in (b) indicate directions of rotation of corotating vortex
clusters.
The blue and red dots in the phase profiles indicate vortices with
clockwise and counterclockwise circulations, respectively.
The field of view is $250 \xi \times 95 \xi$.
See the movie~\cite{movies} showing the dynamics in (a) and (b).
}
\label{f:str}
\end{figure*}
\end{widetext}
First, we consider the case of $a = 0$ in Eq.~(\ref{rpot}), i.e., the
potential moves straight in the $-x$ direction.
It has been known that quantized vortex--antivortex pairs, which we
call vortex dipoles, are released behind an obstacle above a critical
velocity~\cite{Frisch}.
For appropriate values of $v$ and $d$, quantized vortices in the wake
form two kinds of typical patterns~\cite{Sasaki}, as shown in 
Fig.~\ref{f:str}.
In Fig.~\ref{f:str}(a), periodically released vortex dipoles incline
alternately, breaking the symmetry with respect to the $x$ axis.
Since vortex dipoles propagate in the directions as indicated by the
arrows in Fig.~\ref{f:str}(a), the inclined vortex dipoles propagate
obliquely, and the vortex pattern spreads in the downstream.

Figure~\ref{f:str}(b) shows an interesting vortex pattern, which was first
reported in Ref.~\cite{Sasaki}.
The building block of the vortex pattern is not a vortex dipole but a pair
of corotating vortices with the same circulation, which we call a ``vortex 
cluster''.
The vortex clusters with opposite circulations are released alternately,
which form two rows in the wake.
This vortex shedding dynamics and resultant vortex pattern are similar to
those in the BvK vortex street in classical fluids, in that clockwise and
counter clockwise vortices are shed alternately and two rows of vortices
are aligned linearly.
As shown by von K\'arm\'an~\cite{Karman,Lamb}, the vortex pattern as
generated in Fig.~\ref{f:str}(b) is very stable and has a long lifetime.

From the movies~\cite{movies} showing the dynamics in Fig.~\ref{f:str},
it seems that two different instabilities are responsible for the vortex
pattern formations.
The first one is the instability that inclines the vortex dipoles in
Fig.~\ref{f:str}(a).
At the first stage, the vortex dipoles have the symmetry with respect to
the $x$ axis, and after some time the inclination of the vortex dipoles
starts.
The second one is the instability that causes a transition from the vortex
dipole generation to the BvK vortex street generation.
In the movie corresponding to Fig.~\ref{f:str}(b), we can see that the
former changes to the latter dynamically.

For the velocity $v$ slower than the critical velocity of vortex
generation, the flow pattern is a stationary laminar flow.
For a very large velocity, the flow pattern becomes irregular or
a Cherenkov-like pattern~\cite{El,Carusotto}.
The periodic vortex shedding as shown in Fig.~\ref{f:str} is restricted
between these two parameter regions.
(The detailed parameter diagram is obtained in Ref.~\cite{Sasaki}.)
The vortex dipole generation as in Fig.~\ref{f:str}(a) tends to occur for
a smaller obstacle, and the BvK-like vortex street tends to appear for a
larger obstacle.
For an intermediate size of an obstacle, the former (latter) occurs for a
slower (faster) velocity, where the vortex dipole generation changes to
the BvK-like vortex street with an increase in the velocity.

\subsection{Sinusoidal motion of an obstacle}
\label{s:sin}

We next consider the case of sinusoidal trajectories of the obstacle
potential ($a \neq 0$ in Eq.~(\ref{rpot})), and investigate how the
Bvk-like vortex street is affected by the sinusoidal motion.
Since the parameter region for the Bvk-like vortex street to emerge is
narrow~\cite{Sasaki}, we restrict the parameters $V_0$, $d$, and $v$ to
that region, and study the $a$ and $\omega$ dependence of the vortex
shedding dynamics.

\begin{widetext}
\begin{figure*}[tb]
\centerline{%
\includegraphics[width=10cm]{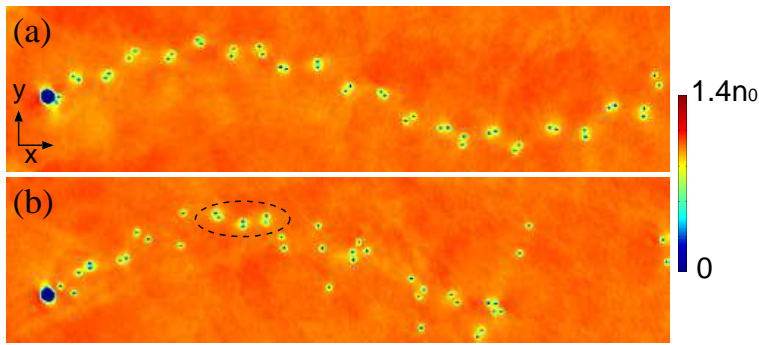}
}
\vskip6pt
\caption{Snapshots of the density profiles for the sinusoidal motion of
the obstacle.
(a) $v = 0.43 v_{\rm s}$, $a = 32 \xi$, and $\omega = 0.006 \tau^{-1}$.
(b) $v = 0.41 v_{\rm s}$, $a = 32 \xi$, and $\omega = 0.008 \tau^{-1}$.
Other parameters are the same as those in Fig.~\ref{f:str}(b).
The field of view is $500 \xi \times 125 \xi$.
}
\label{f:slow}
\end{figure*}
\end{widetext}
First, we examine the low frequency limit, in which $\omega$ is much
smaller than the vortex shedding frequency and the wavelength of the
sinusoidal trajectory is much larger than the distance between vortex
clusters.
The result is shown in Fig.~\ref{f:slow}(a).
The BvK-like vortex street is shed along the smoothly curved trajectory.
For sinusoidal motion given in Eq.~(\ref{rpot}), the minimum and maximum
velocities are $v_{\rm min} = v$ and $v_{\rm max} = (v^2 + a^2
\omega^2)^{1/2}$ and the root-mean-square velocity is $\bar{v} = (v^2 +
a^2 \omega^2 / 2)^{1/2}$.
For the parameters in Fig.~\ref{f:slow}(a), $v_{\rm min}$ and
$v_{\rm max}$ fall into the parameter region of the BvK-like street.
For a larger value of $a \omega$, $v_{\rm min}$, $v_{\rm max}$, or both go
out of the parameter region of the BvK-like street even when $\bar{v}$
falls into that region.
Figure~\ref{f:slow}(b) shows such a case.
The vortex patterns are disturbed, although the BvK-like pattern is partly
observed (dashed circle in Fig.~\ref{f:slow}(b)).

\begin{widetext}
\begin{figure*}[t]
\centerline{%
\includegraphics[width=12cm]{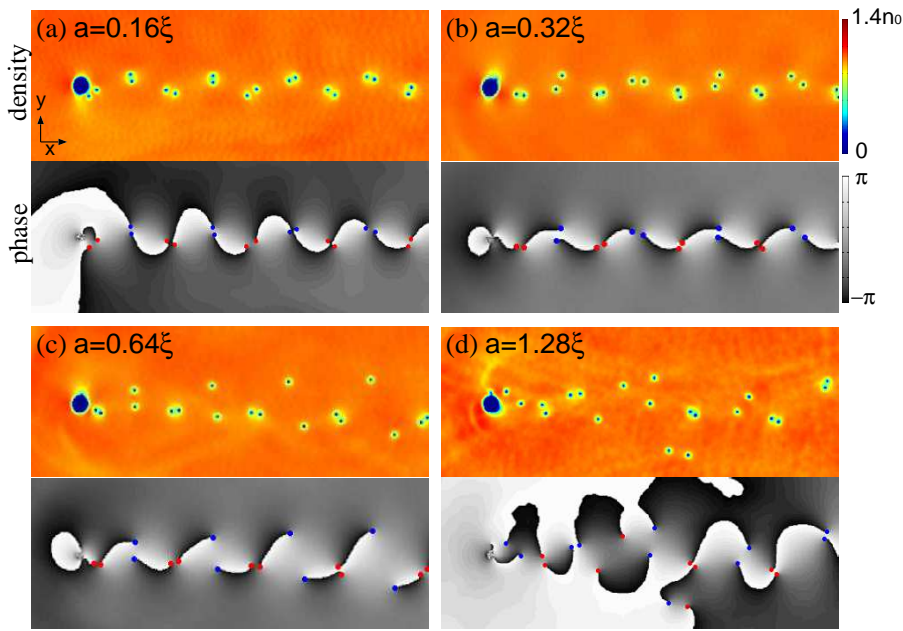}
}
\vskip6pt
\caption{Snapshots of the density ($|\psi|^2$) and phase (arg $\psi$)
profiles for the sinusoidal motion of the obstacle.
The oscillation frequency is $\omega = 0.1 \tau^{-1}$ and the amplitude is
(a) $a = 0.16 \xi$, (b) $a = 0.32 \xi$, (c) $a = 0.64 \xi$, and
(d) $a = 1.28 \xi$.
Other parameters are the same as those in Fig.~\ref{f:str}(b).
The blue and red dots in the phase profiles indicate vortices with
clockwise and counterclockwise circulations, respectively.
The field of view is $250 \xi \times 95 \xi$.
See the movie~\cite{movies} showing the dynamics in (b) and (c).
}
\label{f:sin}
\end{figure*}
\end{widetext}
Figure~\ref{f:sin} shows the case of a larger frequency $\omega = 0.1
\tau^{-1}$, in which the amplitude $a$ is much smaller than that in
Fig.~\ref{f:slow}.
The frequency $\omega = 0.1 \tau^{-1}$ is about twice the shedding
frequency of the vortex clusters.
For a small amplitude, in Fig.~\ref{f:sin}(a), the vortex pattern is
similar to the BvK-like vortex street for straight motion in
Fig.~\ref{f:str}(b).
As $a$ is increased, the symmetry between the two rows of the vortex
clusters is spontaneously broken.
In Fig.~\ref{f:sin}(b), one can see that the vortex clusters in the upper
row is larger than those in the lower row.
In Fig.~\ref{f:sin}(c), the symmetry breaking is significant, where the
vortex clusters in the upper row disintegrate in the downstream.
Whether the vortex clusters in the upper row or lower row become large
depends on the initial condition that breaks the symmetry, since
Eq.~(\ref{GP}) has the symmetry with respect to $y \rightarrow -y$.
In fact, the behaviors of the upper and lower rows can be exchanged by
changing the initial condition (the initial phase of the oscillation in
Eq.~(\ref{rpot}) and added initial noises).
The roles of the upper and lower rows can be exchanged during the
dynamics.
In the movie~\cite{movies} corresponding to Fig.~\ref{f:sin}(b), we can
see that the vortex clusters in the upper row is first enlarged and then
the lower row takes over.
For a larger amplitude $a$, the vortex pattern is disturbed, as shown in
Fig.~\ref{f:sin}(d).
Thus, the degree of the symmetry breaking of the BvK-like vortex street
increases with $a$, until the periodicity is disturbed for $a \gtrsim
\xi$.
It should be noted that the appropriate value of $a$ for the symmetry
breaking phenomenon is $a \sim \xi$, i.e., the amplitude is comparable to
the size of a vortex.
The sinusoidal motion has a significant influence on the vortex shedding
dynamics, even when the amplitude is much smaller than the size of the
obstacle potential.

\begin{widetext}
\begin{figure*}[tb]
\centerline{%
\includegraphics[width=12cm]{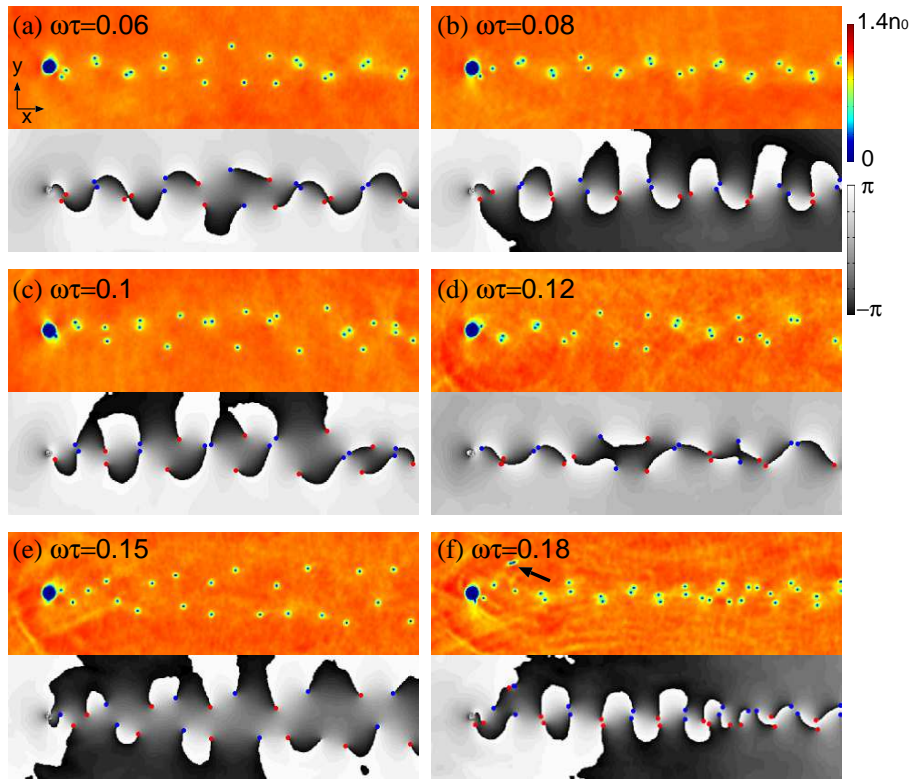}
}
\vskip6pt
\caption{Snapshots of the density ($|\psi|^2$) and phase (arg $\psi$)
profiles for the sinusoidal motion of the obstacle.
The oscillation amplitude is $a = 0.64 \xi$ and the frequency is
(a) $\omega = 0.06 \tau^{-1}$, (b) $\omega = 0.08 \tau^{-1}$,
(c) $\omega = 0.1 \tau^{-1}$, (d) $\omega = 0.12 \tau^{-1}$,
(e) $\omega = 0.15 \tau^{-1}$, and (f) $\omega = 0.18 \tau^{-1}$.
Other parameters are the same as those in Fig.~\ref{f:str}(b).
The blue and red dots in the phase profiles indicate vortices with
clockwise and counterclockwise circulations, respectively.
The field of view is $315 \xi \times 95 \xi$.
}
\label{f:omegad}
\end{figure*}
\end{widetext}
Figure~\ref{f:omegad} shows the $\omega$ dependence of the vortex
dynamics, where the amplitude $a$ is the same as that in
Fig.~\ref{f:sin}(c).
The frequency $\omega$ in Fig.~\ref{f:omegad}(c) is the same as that in
Fig.~\ref{f:sin}, and therefore the parameters in Fig.~\ref{f:omegad}(c)
and Fig.~\ref{f:sin}(c) are identical.
As $\omega$ is decreased, the enlargement of the vortex clusters becomes
intermittent, that is, the vortex shedding as in Fig.~\ref{f:omegad}(c)
and that in Fig.~\ref{f:str}(b) are interchanged intermittently.
At $\omega = 0.08 \tau^{-1}$, an interesting resonance emerges as shown in
Fig.~\ref{f:omegad}(b), where every third vortex cluster is enlarged.
When $\omega$ is further decreased, the enlargement of the vortex clusters
occurs not only in the one row but also in both rows successively, which
is however intermittent, as shown in Fig.~\ref{f:omegad}(a).
For $\omega \lesssim 0.05 \tau^{-1}$, the enlargement of the vortex
clusters ceases and there seems to be no significant effect of the
sinusoidal motion on the BvK-like vortex street (data not shown), that is,
the vortex pattern is similar to that in Fig.~\ref{f:str}(b).

As $\omega$ is increased from $0.1 \tau^{-1}$, the enlargement of the
vortex clusters becomes intermittent, as shown in Fig.~\ref{f:omegad}(d),
which is similar to the vortex pattern in Fig.~\ref{f:omegad}(a).
Around $\omega = 0.15 \tau^{-1}$, the vortex pattern changes drastically,
as shown in Fig.~\ref{f:omegad}(e), where neither vortex dipoles nor
vortex clusters are formed, and the periodically released vortex are
apart from each other.
This may be regarded as that the enlargement of vortex clusters, as shown
in Figs.~\ref{f:omegad}(a) and \ref{f:omegad}(d), occurs successively in
both rows.
For a larger value of $\omega$, the enlargement of vortex clusters becomes
intermittent again, in a manner similar to Figs.~\ref{f:omegad}(a) and
\ref{f:omegad}(d).
The enlargement of vortex clusters ceases and the BvK-like vortex street
is generated for $\omega \gtrsim 0.25 \tau^{-1}$ (data not shown).
For $\omega \simeq 0.18 \tau^{-1}$, extra vortex dipoles are created
during the vortex street generation, as indicated by the arrow in
Fig.~\ref{f:omegad}(f), which may be due to the disturbances by the fast
oscillation of the obstacle.
For $\omega \gtrsim 0.25 \tau^{-1}$, the fast oscillating potential is
effectively averaged in time and the disturbances are decreased.
Thus, when the oscillation frequency of the obstacle is much larger or
smaller than the vortex shedding frequency, the vortex pattern is not much
affected, and when the oscillation frequency $\omega$ is comparable to the
vortex shedding frequency, the vortex pattern depends on $\omega$ in a
complicated manner.
There may be other types of resonance dynamics, which however need fine
tuning of the oscillation frequency.

\begin{figure}[tb]
\centerline{%
\includegraphics[width=7.5cm]{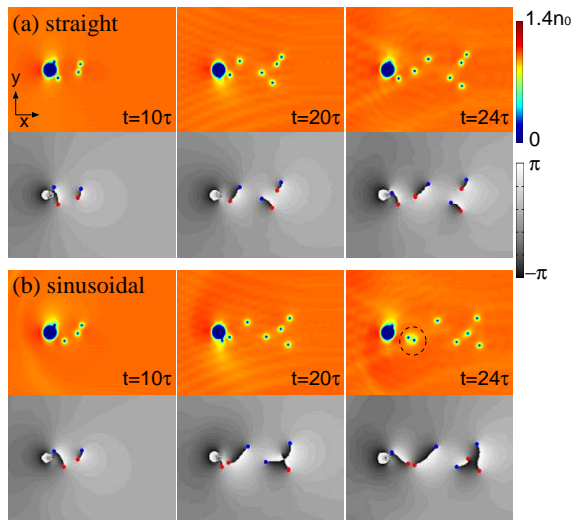}
}
\vskip6pt
\caption{Snapshots of the density ($|\psi|^2$) and phase (arg $\psi$)
profiles.
(a) Without sinusoidal motion.
(b) Sinusoidal motion with $\omega = 0.1 \tau^{-1}$ and $a = 0.32 \xi$.
Other parameters are the same as those in Fig.~\ref{f:str}(b).
The blue and red dots in the phase profiles indicate vortices with
clockwise and counterclockwise circulations, respectively.
The field of view is $125 \xi \times 95 \xi$.
}
\label{f:faster}
\end{figure}
Comparing the movies~\cite{movies} of Figs.~\ref{f:str}(b) and
\ref{f:sin}(b), we find that the BvK-like vortex street emerges faster in
Fig.~\ref{f:sin}(b) than in Fig.~\ref{f:str}(b), i.e., the vibration of
the obstacle seems to enhance the emergence of the BvK-like vortex
street.
Figure~\ref{f:faster} shows the dynamics of the early stage, where the
obstacle starts to move at $t = 0$.
For the straight motion of the obstacle (Fig.~\ref{f:faster}(a)), a vortex
dipoles are first generated ($t = 10 \tau$) and the subsequent vortex
dipoles are inclined.
For the sinusoidal motion of the obstacle (Fig.~\ref{f:faster}(b)), the
sixth and seventh released vortices form a tight vortex cluster (dashed
circle in Fig.~\ref{f:faster}(b)), which is followed by the BvK-like
vortex street with tight vortex clusters.

\section{Conclusions}
\label{s:conc}

We have investigated the dynamics of quantized vortex shedding from an
obstacle moving in a superfluid, in which the obstacle traces a sinusoidal
trajectory.
As in the case of the straight motion of an obstacle, quantized
vortex--antivortex pairs are released when the velocity exceeds some
critical velocity, and for appropriate parameters, a BvK-like vortex
street emerges in the wake.
In contrast to the case of the straight motion, the symmetry between the
two rows of the BvK street is broken; the size of the corotating vortex
clusters is enlarged in either row (Fig.~\ref{f:sin}(c)).
This phenomenon occurs when the oscillation frequency of the obstacle is
close to twice the vortex shedding frequency and the amplitude is
comparable to the healing length.
We have studied the amplitude $a$ and frequency $\omega$ dependence of the
vortex shedding dynamics.
The degree of the symmetry breaking of the BvK-like vortex street
increases with $a$ and the periodic street is destroyed for $a \gtrsim
\xi$ (Fig.~\ref{f:sin}).
The $\omega$ dependence is complicated;
the enlargement of the vortex clusters occurs in various ways, as shown in
Fig.~\ref{f:omegad}.
The sinusoidal motion enhances the symmetry breaking and promotes the
transition from vortex dipole generation to BvK-like vortex street
generation.

The BvK-like vortex street generation in superfluids is an example of a
highly nonlinear phenomenon far from the equilibrium.
When some external vibration is applied to such a system and the frequency
approaches some resonance, we expect that the system exhibits nontrivial
resonant dynamics due to nonlinearity and nonequilibrium.

\begin{acknowledgments}
This work was supported by Grants-in-Aid for Scientific
Research (No.\ 22340116 and No.\ 23540464) from the Ministry of Education,
Culture, Sports, Science and Technology of Japan.
\end{acknowledgments}

\end{document}